\documentclass[pra,preprint]{revtex4}

\usepackage{amsfonts}
\usepackage{amsmath}
\usepackage{epsfig}
\usepackage{theorem}

\def\tr{{\rm tr}}
\def\ket#1{|#1\rangle}
\def\bra#1{\langle#1|}
\def\Lindblad#1{2a #1 a^{\dag} -a^{\dag}a #1 - #1 a^{\dag}a }
\newcommand{\one}{\mbox{\tt 1}\hspace{-0.057 in}\mbox{\tt l}}

\begin{document}

\title{Information dynamics in cavity QED}

\author{Andrei N. Soklakov}
\author{R\"udiger Schack}
\affiliation{Department of Mathematics,
Royal Holloway, University of London,
Egham, Surrey TW20 0EX, UK}

\date{December 2, 2002}

\begin{abstract}
  A common experimental setup in cavity quantum electrodynamics (QED) consists
  of a single two-level atom interacting with a single mode of the
  electromagnetic field inside an optical cavity. The cavity is externally
  driven and the output is continuously monitored via homodyne measurements.
  We derive formulas for the optimal rates at which these measurements provide
  information about (i) the quantum state of the system composed of atom and
  electromagnetic field, and (ii) the coupling strength between atom and
  field. We find that the two information rates are anticorrelated.
\end{abstract}

\maketitle

\section{Introduction}

In this paper, we consider a system consisting of a single two-level
atom, located inside an externally driven optical cavity. 
The atom interacts with a single light mode inside the cavity, 
which is coupled to the environment via a partially transparent 
mirror. The output field from the cavity is
monitored using the homodyne detection scheme, 
in which the cavity output is
added to a reference field at a beam-splitter and then analyzed by two
photodetectors (see Fig.~{\ref{Setup}}).

In two recent ground-breaking 
experiments~\cite{HoodEtAl_1998,PinkseEtAl_2000}, 
a similar setup was used to observe 
the trajectory of the atom inside the cavity.
In these experiments,
the atomic position was inferred from the strength of the 
atom-cavity coupling, which can be estimated directly from the 
photocurrents~\cite{Mabuchi_1996}. 
To characterize the performance of this {\it atom-cavity
microscope}, Kimble introduced a quantity called 
{\it optical information}~\cite{HoodEtAl_2000,DohertyEtAl_2001},
 which measures the rate at which the measurement provides 
information
about the system. In 
Refs.~\cite{HoodEtAl_2000,DohertyEtAl_2001}, however, no formal
definition of optical information is given, and only a heuristic derivation
of its value is provided. Recently, a number of related definitions of
information gain have been proposed  and investigated numerically for a
simpler quantum-optical system~\cite{Gambetta2001}.

The question of how much information about a monitored system is provided by a
continuous measurement is interesting in its own right. In this paper, we
consider two, as it turns out, complementary cases. In Sec.~\ref{sec:state}
we calculate the optimal rate at which a homodyne measurement provides
information about the quantum state of the system composed of the
electromagnetic field and the internal state of the atom. In
Sec.~\ref{sec:coupling}, we calculate the optimal rate at which the
measurement gives information about the coupling
strength between the atom and the intra-cavity field.
In Sec.~\ref{CavityQED}, we introduce our mathematical model and main
assumptions, and Sec.~\ref{sec:discussion} concludes the paper with a short
discussion.

\section{Model} \label{CavityQED}

The evolution of the state, $\rho$, of an open quantum system subject to
a continuous measurement can often be described by a stochastic master
equation of the 
form~\cite{BelavkinStaszewski_1992,WisemanMilburn_1993b,WisemanMilburn_1993a}
\begin{equation}                                           \label{StochasticME}
d\rho={\cal L}(\rho)dt+{\cal M}(\rho)dW\,,
\end{equation}
which is understood in the sense of the It\^o stochastic differential
calculus~\cite{Gardiner_1985b}.
Any particular measurement record is represented
by some realization of the stochastic process $W(t)$.
The superoperator ${\cal L}$ is linear and
defines the ``unconditional'' evolution, i.e., the evolution
in the absence of measurements. By contrast,
the superoperator ${\cal M}$ is nonlinear and
accounts for the effects of the measurement.

\begin{figure} 
\begin{center}
\epsfig{file=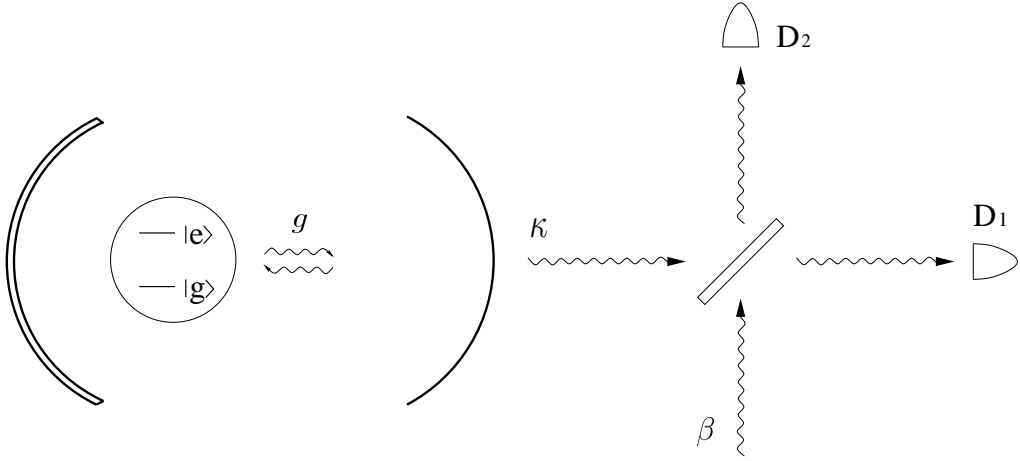,width=13.5cm}
\end{center}
\caption{Homodyne or heterodyne measurements in cavity QED.
See the text for an explanation of the parameters.
}
\label{Setup}
\end{figure}

The cavity-QED system we are considering here is illustrated in
Fig.~{\ref{Setup}}.  A single two-level atom, with ground state $|{\rm
  g}\rangle$ and excited state $|{\rm e}\rangle$ is located inside a
high-finesse optical cavity, which is driven by an external laser field of
strength $E$. The atom interacts with a single light mode of the cavity.
We denote by $g$ the
strength of the atom-cavity coupling, and by $\kappa$ the cavity field
decay rate.
In this paper, we assume that the
atom, the cavity and the driving laser are all resonant.  

In the absence of measurements, ${\cal M}=0$ in Eq.~(\ref{StochasticME}), 
and the joint density operator 
of atom and intra-cavity field, $\rho$, obeys the master equation
\begin{equation}                                     \label{OnlyStandardApprox}
\dot{\rho}={\cal L}(\rho) \;,
\end{equation}
where 
\begin{equation}                                      \label{L}
{\cal L}(\rho) = \big[ E(a^\dag -a)
      +g(a^\dag\sigma-\sigma^\dag a) ,\rho\,\big]
         +\kappa( \Lindblad{\rho} )\,.
\end{equation}
Here $a$ is the annihilation operator for the cavity field mode, and 
$\sigma = |{\rm g}\rangle\langle{\rm e}|$.
In the following we focus our attention on the important case of
the strong driving regime ($E/g\gg 1$).
In this regime the system approaches a steady
state of the form~\cite{AlsingCarmichael_1991}
\begin{equation}                                                                   \label{RhoSS1}
\rho_{\rm ss}^\alpha=\frac{1}{2}\big( \ket{\alpha;+}\bra{\alpha;+}
                      +\ket{\alpha^*;-}\bra{\alpha^*;-}\big)\;.
\end{equation}
Here $\ket{\alpha;+}$
and $\ket{\alpha^*;-}$ are two orthogonal
quantum states defined by
\begin{eqnarray}                                                                       \label{basis}
\ket{\alpha;+}&=&\frac{1}{\sqrt{2}}
                 \ket{\alpha}\big(\ket{{\rm g}}+i\ket{{\rm e}}\big)\,,\cr
\ket{\alpha^*;-}&=&\frac{1}{\sqrt{2}}\ket{\alpha^*}\big(\ket{{\rm g}}
                                 -i\ket{{\rm e}}\big)\,,
\end{eqnarray}
where $\ket{\alpha}$ is the coherent field state with amplitude
\begin{equation}                                                                      \label{alpha}
\alpha=
\frac{E}{\kappa}\left[1-\Big(\frac{g}{2E}\Big)^2
+i\frac{g}{2E}\sqrt{1-\Big(\frac{g}{2E}\Big)^2}\;\right]\,.
\end{equation}

Let us now assume that the field escaping from our system
undergoes a continuous measurement via the standard
homodyne measurement scheme~\cite{PlenioKnight_1998}. 
In this scheme we have only one free complex parameter:
the reference field $\beta$ that is added to the cavity output at
a beam splitter prior to the detection by two detectors $D_1$ and $D_2$ (see
Fig.~{\ref{Setup}}). 
If the measurement record consists of the scaled difference photocurrent
$dq/dt\equiv(I_2-I_1)/|\beta|$, where $I_1$ and $I_2$ are the photocurrents
detected by $D_1$ and $D_2$, respectively, then
the measurement operator ${\cal M}$ in the stochastic master
equation~(\ref{StochasticME}) is given 
by~\cite{WisemanMilburn_1993b,WisemanMilburn_1993a}
\begin{equation}                                           \label{M_rho_c}
{\cal M}(\rho) = 
     \sqrt{2\kappa\eta}\,
              \Big(e^{-i\phi}a\rho
                         +e^{i\phi} \rho a^{\dag}
                 -\tr\!\left[\rho(e^{-i\phi}a
                 +e^{i\phi}a^{\dag})\right]
           \rho
              \Big) \;,
\end{equation}
where $\eta$ is the efficiency of the photodetection and $\phi=\arg \beta$
is the phase of the reference field. 
The realization of the Wiener process $W(t)$ in Eq.~(\ref{StochasticME}) is
connected to the experimentally observed difference photocurrent
via the relation~\cite{WisemanMilburn_1993b,WisemanMilburn_1993a}
\begin{equation}                                                \label{DiffPhotocurrent}
dq=
2\kappa\eta\;\tr\!\left[\rho(e^{i\phi}a^\dag
                                                                                     +e^{-i\phi} a)
                                               \right] dt
                  +\sqrt{2\kappa\eta}\,dW\;.
\end{equation}
In the special case $\rho=\rho_{\rm ss}^\alpha$ this
equation becomes
\begin{equation}                                \label{DiffPhotocurrent2}
dq
=4\kappa\eta\,{\rm Re}(\alpha)\cos\phi\; dt
                  +\sqrt{2\kappa\eta}\,dW\;.
\end{equation}

\section{Inference} \label{DeltaHSection}

\subsection{Information about the quantum state}   \label{sec:state}

The amount of information provided by the measurement 
about the quantum state, $\rho$,
is quantified by the reduction of the von~Neumann entropy,
\begin{equation}
H(\rho)\equiv-\tr(\rho\ln\rho)\,.   \label{vonNeumannEntropy}
\end{equation}
In this subsection we calculate the rate at which this entropy
changes as a result of measurements. 
The average rate of entropy change in the presence of continuous
observations is given by 
\begin{equation}                      \label{defDotH}
\langle\dot{H}(\rho)\rangle
= \lim_{\Delta t\to 0}
\left\langle
\frac{H(\rho+\delta)-H(\rho)}{\Delta t}
\right\rangle
\end{equation}
where the average $\langle\cdot\rangle$
is taken over all possible measurement outcomes
observed over the time $\Delta t$, and 
\begin{equation}                                         \label{deltaDef}
\delta={\cal L}(\rho)\Delta t + {\cal M}(\rho) \Delta W\,.
\end{equation}

We now assume that the 
system is initially in a steady state, 
$\rho_{\rm ss}$, defined by the relation 
\begin{equation}
{\cal L}(\rho_{\rm ss})=0 \;.   \label{eq:steady}
\end{equation}
For simplicity, we furthermore assume that  
\begin{equation}
[{\cal M}(\rho_{\rm ss}),\rho_{\rm ss}]=0 \;.   \label{eq:commute}
\end{equation} 
This last assumption
is satisfied by Eqs.~(\ref{RhoSS1}) and (\ref{M_rho_c}).
At the end of this section
we will outline a more general framework which is free from these 
assumptions.
Since the stochastic master equation (\ref{StochasticME}) preserves
the positivity of the density operator, $\rho$, 
the support ${\cal M(\rho_{\rm ss})}$ must lie in the support of 
$\rho_{\rm ss}$, ${\rm supp} ({\cal M(\rho_{\rm ss})})\subseteq 
{\rm supp}(\rho_{\rm ss})$. 
Now consider a basis in which $\rho_{\rm ss}$ and 
${\cal M}(\rho_{\rm ss})$ are both diagonal. 
In this basis, let $a_1,\dots,a_m$ be the nonzero diagonal 
elements of $\rho_{\rm ss}$ and let $b_1,\dots,b_m$ be
the corresponding diagonal elements of $\cal{M}(\rho_{\rm ss})$.

Substituting $\rho=\rho_{\rm ss}$ into Eq.~(\ref{defDotH}), we obtain
\begin{eqnarray}                             \label{AtTheSteadyState}                            
\langle\dot{H}(\rho_{\rm ss})\rangle
&=&\lim_{\Delta t\to 0}\left\langle\frac{ 
H(\rho_{\rm ss}+{\cal M}(\rho_{\rm ss})\Delta W)
-H(\rho_{\rm ss})}{\Delta t}
\right\rangle\cr
&=&
\lim_{\Delta t\to 0}\left\langle\frac{ -
\sum_{k=1}^{m}(a_k+b_k\Delta W)\ln (a_k+b_k\Delta W)+\sum_{k=1}^{m}
a_k\ln a_k
}{\Delta t}
\right\rangle\cr
&=&-\sum_{k=1}^{m}\frac{(b_k)^2}{2a_k}\,,
\end{eqnarray}
where we have expanded the logarithm to second order in $\Delta W$
and used the relation $\langle(\Delta W)^2\rangle=\Delta t$.

For any state $\rho_{\rm ss}^\alpha$ of the form~(\ref{RhoSS1}) we have
\begin{equation}                                             \label{dRhoCase1}
 {\cal M}(\rho_{\rm ss}^\alpha)=
\sin\phi\,
\sqrt{2\kappa\eta}\, {\rm Im}(\alpha)\Big{(}
                                       |\alpha;+\rangle\langle\alpha;+|
                                    -   |\alpha^*;-\rangle\langle\alpha^*;-|
                                        \Big{)}\,.
\end{equation}
Since $\rho_{\rm ss}^\alpha$ and ${\cal M}(\rho_{\rm ss}^\alpha)$ commute, we
can apply Eq.~(\ref{AtTheSteadyState}). For the particular value for $\alpha$
given by Eq.~(\ref{alpha}), the diagonal elements of 
$\rho_{\rm ss}^\alpha$ and ${\cal M}(\rho_{\rm ss}^\alpha)$ are
\begin{equation}                                \label{dRhoCase11}
a_{1,2}=1/2\,,\ \ \ {\rm and} \ \ \   
b_{1,2}=\pm
g\,\sin\phi\,
\sqrt{\frac{\eta}{2\kappa}\left(1-(\frac{g}{2E})^2\right)}\,.
\end{equation}
Finally, we obtain the main result of this subsection:
the rate of information gain about the quantum state 
of the system is
\begin{equation}   \label{eq:RQ}
 R_Q  \equiv -\langle\dot{H}(\rho_{\rm ss}^\alpha)\rangle = 
    \frac{g^2\eta}{\kappa}
                  \left(1-\big(\frac{g}{2E}\big)^2\right)\sin^2\phi\,.
\end{equation}
Here the minus sign indicates that the information gain corresponds
to the reduction of uncertainty about the system state which is measured 
by the von~Neumann entropy.

We now finish this subsection by
developing some general formalism that makes no assumptions regarding the
superoperators ${\cal L}$ and ${\cal M}$.  Equation~(\ref{vonNeumannEntropy})
can be expanded in the form
\begin{eqnarray}                                                                   \label{defH}
H(\rho)
&=&\tr\Big[\sum_{n=1}^{\infty}
                         \frac{(-1)^n}{n}\rho(\rho-\one)^n\Big]\cr
&=&\tr\Big[\sum_{n=1}^{\infty}
                         \frac{(-1)^n}{n}
                                  (\tilde{\rho}^{n+1} + \tilde{\rho}^n)\Big]\,,
\end{eqnarray}
where $\tilde{\rho}=\rho-\one$.
Equation~(\ref{defDotH}) then becomes
\begin{equation}                                            \label{dotHaveraged}
\langle\dot{H}(\rho)\rangle
=\lim_{\Delta t\to 0} \sum_{n=1}^{\infty}\frac{(-1)^n}{n\,\Delta t}
         \left\langle
             \tr[(\tilde{\rho}+\delta)^{n+1}-\tilde{\rho}^{n+1}]
           +\tr[(\tilde{\rho}+\delta)^n -\tilde{\rho}^n]
          \right\rangle \;.
\end{equation}
Keeping the terms to the second order in $\delta$, we have
\begin{equation}                        \label{GeneralBinom}
(\rho+\delta)^n
=\rho^n
   +\sum_{l=0}^{n-1}
               \rho^l\;
               \delta\;
               \rho^{n-1-l}
   +\sum_{p=0}^{n-2}\sum_{q=0}^{n-2-p}
                                                  \rho^p\;
                                                  \delta\,\;
                                                  \rho^{n-2-(p+q)}
                                                  \delta\;
                                                  \rho^q
   +O(\delta^3)\,.
\end{equation}
Using the cyclic property of the trace,
\begin{eqnarray}                 \label{TrRPDnMinusRn}
\tr[(\rho+\delta)^n-\rho^n]
&=&\tr\Big[
   n\rho^{n-1}\;\delta
   +\sum_{p=0}^{n-2}\sum_{q=0}^{n-2-p}
                                                  \rho^{p+q}\;
                                                  \delta\,\;
                                                  \rho^{n-2-(p+q)}
                                                  \delta
   +O(\delta^3) \Big]   \cr
&=&\tr\Big[ n\rho^{n-1}\;\delta
                  +\sum_{s=0}^{n-2}
                  (s+1)
                  \rho^{s}\;
                  \delta\,\;
                  \rho^{n-2-s}
                  \delta
   +O(\delta^3) \Big]\,.
\end{eqnarray}
Combining Eqs.~(\ref{dotHaveraged}) and (\ref{TrRPDnMinusRn}), we obtain
\begin{align}
\langle\dot{H}(\rho)\rangle
=
\lim_{\Delta t\to 0} \sum_{n=1}^{\infty}\frac{(-1)^n}{n\,\Delta t}
         \tr\Big{\langle}\,
           (n+1)\tilde{\rho}^n\delta
           +&\sum_{s=0}^{n-1}(s+1)\tilde{\rho}^s\delta
                                                        \tilde{\rho}^{n-1-s}\delta  \cr
           +n\tilde{\rho}^{n-1}\delta
                 +&\sum_{s=0}^{n-2}(s+1)\tilde{\rho}^s\delta
                                                        \tilde{\rho}^{n-2-s}\delta \,
          \Big{\rangle}\,.
\end{align}
Substituting Eq.~(\ref{deltaDef}) and using
$\langle\Delta W\rangle=0$,
$\langle(\Delta W)^2\rangle=\Delta t$
we finally arrive at
\begin{align}                             \label{dotHafterSubst_deltaDef}
\langle\dot{H}(\rho)\rangle
=
 \sum_{n=1}^{\infty}\frac{(-1)^n}{n}
         \tr\Big{(}\,
           &[(n+1)\tilde{\rho}^n
                +n\tilde{\rho}^{n-1}]{\cal L}(\rho)\cr
           &+\sum_{s=0}^{n-1}
                           (s+1)\tilde{\rho}^s{\cal M}(\rho)
                           \tilde{\rho}^{n-1-s}{\cal M}(\rho)\cr
           & +\sum_{s=0}^{n-2}
                            (s+1)\tilde{\rho}^s{\cal M}(\rho)
                            \tilde{\rho}^{n-2-s}{\cal M}(\rho) \,
          \Big{)}\,.
\end{align}
This formula is valid without any restrictions on the superoperators ${\cal
  L}$ and ${\cal M}$. It is straightforward to show that, under the previous
assumptions $\rho=\rho_{\rm ss}$ and $[{\cal M}(\rho_{\rm ss}),\rho_{\rm
  ss}]=0$, Eq.~(\ref{dotHafterSubst_deltaDef}) reduces to
  Eq.~(\ref{AtTheSteadyState}) as required.

\subsection{Information about the atom-cavity coupling}  \label{sec:coupling}

Since the quantized degrees of freedom described by the density operator
$\rho$ of the previous sections do not include the atomic position, a
different approach is needed to obtain information about the atom's trajectory
inside the cavity. The key is the fact that the atom-cavity coupling $g$
depends on the position of the atom in a known way (see, e.g.,
Ref.~\cite{HoodEtAl_2000}). In this subsection, we will therefore focus on
calculating the rate at which the measurement provides information about the
parameter $g$.

Let $q$ be a photocharge obtained by integrating the
difference photocurrent, given by
Eq.~(\ref{DiffPhotocurrent}), over a small time interval $\Delta t$.
Introduce a small parameter $\epsilon$ such that
$\kappa\Delta t \sim \epsilon^{3/2}$ and
$|\beta|\sim \epsilon^{-1}$.
The existence of such $\epsilon$ is assumed in the
standard derivation of Eqs.~(\ref{M_rho_c})
and (\ref{DiffPhotocurrent}) \cite{WisemanMilburn_1993a, WisemanMilburn_1993b}.
Equation (\ref{DiffPhotocurrent}) is derived in the limit of small
$\epsilon$ when $q$ can be treated as
a Gaussian random variable, 
\begin{equation}
G(q,\langle q \rangle,v^2)\equiv \frac{1}{\sqrt{2\pi v^2}}
\exp\Big{(}-\frac{(q-\langle q \rangle)^2}{2v^2}\Big{)}\,,
\end{equation}
with mean
\begin{equation} \label{q_average}
\langle q \rangle=2\kappa\eta\,
                                  \tr[\rho(e^{i\phi}a^\dag+e^{-i\phi}a)]
                                  \Delta t +O(\epsilon^2)
\end{equation}
and variance
\begin{equation}
v^2= 2\kappa \eta \Delta t +O(\epsilon^2)\,.
\end{equation}
So far we have
treated the atom-cavity coupling $g$ as a known parameter.
This means that
the above equations give us the conditional probability density
$P(q|g)$ of registering measurement result $q$
given that the atom-cavity coupling is equal to $g$,
\begin{equation} \label{Pqg}
P(q|g)=G(q,\langle q \rangle,v^2).
\end{equation}
The conditional probability of the atom-cavity coupling to be equal to $g$
given a particular measurement result $q$ can be derived
from the Bayes rule, 
\begin{equation} \label{Pgq}
P(g|q)=\frac{P(q|g)P(g)}{\int P(q|g)P(g) dg}\,,
\end{equation}
where $P(g)$ is a probability distribution that characterizes
our knowledge of $g$ prior to obtaining $q$.

To quantify the information gain, we use the entropy $S$ of a continuous
probability distribution $f(x)$ defined by
\begin{equation}
S[f(x)]\equiv -\int f(x)\ln f(x)\, dx\,.
\end{equation}
The average rate at which observation of $q$ gives us information 
about $g$ is then given by
\begin{equation}                  \label{eq:defRg}
R_g\equiv\lim_{\Delta t\to 0}
   \Big\langle\frac{-\Delta S}{\Delta t}\Big\rangle
=  \lim_{\Delta t\to 0}{1\over\Delta t} \int (-\Delta S) P(q|g)\, dq \;,
\end{equation}
where 
\begin{equation}
\Delta S =  S[P(g|q)] - S[P(g)] \;.
\end{equation}

On the relatively slow time scales of the atomic motion one can
assume to a very good approximation that the quantized 
degrees of freedom described by the density matrix $\rho$
are in the steady state given by Eq.~(\ref{RhoSS1}).
In this case Eq.~(\ref{q_average}) gives
\begin{eqnarray}        \label{eq:quick}
(q-\langle q \rangle)^2 &=& q^2-4q\kappa\eta\,
\tr\big[\rho(e^{i\phi}a^\dag+e^{-i\phi}a)\big]
                                  \Delta t +O(q\epsilon^2)+O(\epsilon^3) \cr
 &=& q^2-8qE\eta  \cos\phi\,\Big(1-({g\over2E})^2\Big) \Delta t 
  +O(q\epsilon^2)+O(\epsilon^3) \;.
\end{eqnarray}
We can now derive a simple formula for $R_g$ if we make the further convenient
assumption that the prior distribution $P(g)$ is a Gaussian with variance
$v_0^2$,
\begin{equation} \label{Pg}
P(g)=G(g,\langle g \rangle, v_0^2)\,.
\end{equation}
The posterior probability density $P(g|q)$ will then also be
a Gaussian, 
\begin{equation}
P(g|q)=G(g,m,v_1^2) \;,
\end{equation}
where the variance $v_1^2$ is given by 
\begin{equation}
v_1^2 = \frac{v_0^2\kappa E}{\kappa E+  q v_0^2 \cos\phi}\,.
\end{equation}
This expression, where we neglected the $O(q\epsilon^2)$ and $O(\epsilon^3)$
terms in Eq.~(\ref{eq:quick}), is valid for small $q$.
The value of the mean $m$ is irrelevant for the current argument.
By a direct calculation, we obtain the entropy change
\begin{equation}
\Delta S = -{1\over2}\ln(v_0^2/v_1^2) = -{1\over2}\ln
  \big( 1 + q{\cos\phi\over\kappa E}v_0^2\big) \simeq
- q{\cos\phi\over2\kappa E}v_0^2 \;.
\end{equation}
Due to the limit $\Delta t\to0$, the Gaussian $P(q|g)$ in Eq.~(\ref{eq:defRg})
is strongly peaked at small values of $q$. Hence we can substitute the above
estimation of $\Delta S$ into Eq.~(\ref{eq:defRg}). We thus obtain our final
result, the average information rate
\begin{equation} \label{dIg}
R_g=\frac{2v_0^2\eta}{\kappa}
  \Big( 1- \big({g\over2E}\big)^2\Big)\cos^2\phi \;.
\end{equation}

\section{Discussion}   \label{sec:discussion}

Comparing the expressions Eq.~(\ref{eq:RQ}) for the information rate $R_Q$ and
Eq.~(\ref{dIg}) for the information rate $R_g$, one notices their very similar
structure. The rate $R_Q$, referring to information about the quantum state
of the atom-cavity system, is proportional to $\sin^2\phi$, whereas the rate
$R_g$, referring to information about the parameter $g$, is proportional to
$\cos^2\phi$. 
In both formulas, $\phi$ is the phase of the reference field $\beta$.
The proportionality factor in $R_g$ is obtained from the proportionality
factor in $R_Q$ by substituting for $g^2$ twice the prior variance, $2v_0^2$, 
of the random variable $g$.

This means that there is a trade-off between the two information rates: the
more we learn about the coupling parameter (and hence the atomic position),
the less we can learn about the atom-cavity quantum state and vice versa.
This is a manifestation of the uncertainty principle applied to the conjugate
field quadratures $X=a+a^\dag$ and $Y=i(a-a^\dag)$. Let us elaborate on this
point by considering our starting equations,
i.e.~Eq.~(\ref{DiffPhotocurrent}), which gives our measurement record, and
Eqs.~(\ref{StochasticME}), (\ref{L}), and (\ref{M_rho_c})
for the corresponding change of the system density
matrix. As before, we assume the steady state $\rho_{\rm ss}$ as the initial
condition for these equations. The qualitative conclusions of this
discussion do not depend on any specific prior distribution for $g$.

We notice that the average photocurrent is proportional to the expectation
value of the expression $e^{i\phi}a^\dag+e^{-i\phi}a$.  In this sense the
cases $\phi=0$ and $\phi=\pi/2$ correspond to measuring the expectation values
of $X$ and $Y$ quadratures respectively. In the case of $\phi=0$ the nonlinear
contribution from the measurement, given by Eq.~(\ref{M_rho_c}), 
becomes proportional to
$\rho_{\rm ss}^\alpha$. From Eq.~(\ref{StochasticME}) we therefore see that the
system density matrix remains equal to $\rho_{\rm ss}^\alpha$ at all times,
independent of the measurement record. This means that for $\phi=0$ the
measurement record contains no information about the quantum state of the
system, in agreement with~(\ref{eq:RQ}). At the same time the setting
$\phi=0$ maximizes the average photocurrent,
which carries information about $g$ [see Eq.~(\ref{DiffPhotocurrent2})].
The second stochastic term in (\ref{DiffPhotocurrent2})
does not depend on $g$ and so, from the point of view of learning
about $g$, this stochastic term is nothing but noise
superimposed on the average photocurrent. 
We therefore see that maximizing the average photocurrent 
increases the
signal-to-noise ratio for measuring $g$ and thus $\phi=0$ gives us the maximum
information about $g$, in agreement with Eq.~(\ref{dIg}).  In summary, the
case $\phi=0$ is ideal for measuring $g$ and it gives no information about the
internal state.  In the case $\phi=\pi/2$ we have the opposite situation: we
obtain maximum information about the atom-cavity state and no information about
$g$, because the measured photocurrent becomes independent of $g$.
For a numerical comparison of the two information rates for a simpler
quantum-optical system and a wide range of detection schemes, 
see Ref.~\cite{Gambetta2001}.

At first sight, learning about the quantum state of the system seems to be a
very different problem form learning about the atomic position.  It so happens
that within our framework these two tasks are best accomplished by monitoring
conjugate observables, namely the $X$ and $Y$ field quadratures.  With a
straightforward modification our calculations can be also applied to the
heterodyne detection scheme. For heterodyne detection, both field quadratures
are monitored simultaneously, at the cost of reducing the respective
maximal information rates.

\section{Acknowledgments}

ANS would like to thank Charlene Ahn, 
Kevin Birnbaum, Jeff Kimble, and Theresa Lynn for many
helpful discussions. Part of this work was carried out at the Institute for
Quantum Information at Caltech, the hospitality of which is gratefully
acknowledged. 
This research was partially funded by project Q-ACTA of the IST-FET
programme of the EU.


\end{document}